# Interpersonalizing Intimate Museum Experiences

Karin Ryding[a], Jocelyn Spence[b], Anders Sundnes Løvlie[a*], Steve Benford[b]

[a]*Digital Design Department, IT University of Copenhagen, Copenhagen, Denmark*

[b]*Mixed Reality Lab, The University of Nottingham, Nottingham, UK*

*Corresponding author: Anders Sundnes Løvlie, IT University of Copenhagen, Rued Langgaards Vej 7, 2300 Copenhagen S, Denmark. Email: asun@itu.dk. ORCID: https://orcid.org/0000-0003-0484-4668

# Interpersonalizing Intimate Museum Experiences

We reflect on two museum visiting experiences that adopted the strategy of interpersonalization in which one visitor creates an experience for another. In the *Gift* app, visitors create personal mini-tours for specific others. In *Never let me go*, one visitor controls the experience of another by sending them remote instructions as they follow them around the museum. By reflecting on the design of these experiences and their deployment in museums we show how interpersonalization can deliver engaging social visits in which visitors make their own interpretations. We contrast the approach to previous research in customization and algorithmic personalization. We reveal how these experiences relied on intimacy between pairs of visitors but also between visitors and the museum. We propose that interpersonalization requires museums to step-back to make space for interpretation, but that this then raises the challenge of how to reintroduce the museum's own perspective. Finally, we articulate strategies and challenges for applying this approach.

**Introduction**

A powerful aspect of digital technologies is their ability to personalize the user experience, for example by capturing data about people's preferences and behaviours, developing algorithms to profile them, and then adapting their interactions accordingly. Personalization has been applied across a wide variety of domains including information retrieval and hypermedia (Steichen et al., 2012), social media (Abdel-Hafez & Xu, 2013), online learning (Sunar et al., 2016), games (Karpinskyj et al., 2014) and museum visiting (Kontiza et al., 2018; Kuflik et al., 2011; Jonathan Lee & Paddon, 2017; van Tuijn & et al, 2016) which is the focus of this paper. In turn, HCI has long been concerned with how users set about customizing their own experiences for themselves, from software productivity tools (Mackay,

1991) to games (Dyck et al., 2003) and many examples of hybrid physical-digital experiences (Ames et al., 2014; Benford et al., 2018; Cheatle & Jackson, 2015; Rosner & Ryokai, 2009; Tsaknaki et al., 2014).

In this article we explore a novel approach to personalization called *interpersonalization* (Eklund, 2020) in which one human personalizes the experience of another. We explore this idea in the context of museum visiting, considering how it can help museums respond to the need to diversify audiences, deliver meaningful experiences to individuals and incorporate digital technologies into hybrid experiences (Falk & Dierking, 2012; Parry, 2010; Simon, 2010). We introduce two different ways in which one visitor can personalize the experience of another, based upon two distinct aspects of social interaction. The first, an embodied experience scaffolded by the *Gift* app, builds upon the powerful social transaction of gifting and involves one visitor making an individual mini-tour as a gift for another. The second, called *Never let me go*, draws on social play with one visitor vicariously controlling the experience of another as they follow them around the museum. We report on a Research Through Design process (Zimmerman et al., 2007) with a significant element of Research in the Wild (Rogers & Marshall, 2017) in which we trialled and studied these two different experiences with the public in museums.

We reflect across these examples and draw out three key themes. First, we describe interpersonalization as a format of interaction, noting that the personal relationship between the visitors brings them a new perspective on the museum exhibits, helping them see new meaning and relevance in the artefacts in light of their personal connection. Second, we suggest that our designs rely on establishing intimacy, both between the visitors but also in terms of visitors' relationships to the museum and its exhibits. Finally, we discuss how the museum can consider its own role in such experiences, especially stepping back to make space for visitors' own interpretations (Sengers & Gaver, 2006).

We suggest that these kinds of interpersonalized and intimate visiting experiences can bring value to museums as they represent an approach to creating new museum experiences that is both powerful, drawing on the strength of the visitors' personal relations, but also lightweight with regard to required technical infrastructure, both of our examples being deployed as public web apps. Moreover, our findings suggest that they may lead to new possibilities for inspiring visitors to share their own interpretations of museum content, which may be valuable for facilitating increased visitor engagement.

With a wider perspective, we argue that interpersonalization may speak to how we understand personalization of experiences elsewhere. Interpersonalization is focused on the interpersonal relation, and as such is qualitatively different from personalization/ customization when understood as adapting to individual users – "individualization" (Bowen & Filippini-Fantoni, 2004). At the same time, these experiences are also distinct from the large scale, public sharing of experiences that (typically) takes place on social media. Placed on a level between the individual and the (massively) social, but simultaneously engaging with the museum as a site of public learning and discovery, they offer one-on-one experiences that blend qualities of the intimate with those of the public.

**Related work**

Our consideration of related work spans two core themes of our paper, interpersonalization and intimacy, with particular attention to the relevance of each to both HCI and interactive museum experiences.

*Interpersonalization*

The suppliers of products, including goods, services and interactive experiences, have long been concerned with how to differentiate them, that is how to adapt them to different consumers so as to add value. There have been two broad strategies for achieving this.

The first, *customization*, is typically viewed as a human-driven, manual process in which either the supplier or consumer (or perhaps both) adapt the product (Jay Lee et al., 2015). HCI has previously considered how users customize various digital products for themselves, including workplace productivity tools (Mackay, 1991), computer games (Dyck et al., 2003) and crafting and making, which typically involve elements of both physical and digital customization (Ames et al., 2014; Benford et al., 2018; Cheatle & Jackson, 2015; Rosner & Ryokai, 2009; Tsaknaki et al., 2014). Others have described how users come to appropriate technologies for their own purposes during the course of practice (Dourish, 2001) so as to support situatedness, dynamics and ownership (Dix, 2007) and discussed how processes of appropriation, non-appropriation and disappropriation may result in technologies being designed, in-use and ultimately rejected (Carroll et al., 2001).

The second strategy, *personalization*, refers to a largely automated process in which algorithms draw on personal data to adapt products for consumers (Arora et al., 2008; Sundar & Marathe, 2010). Algorithmic personalization has also been widely applied to digital technologies, including to information retrieval and hypermedia (Steichen et al., 2012), social media (Abdel-Hafez & Xu, 2013), online learning (Sunar et al., 2016) and games (Karpinskyj et al., 2014). In discussing personalized information retrieval, Ghorab et al. (2013) distinguish between approaches that are individualized (operate at the levels of the individual), community-based (operate at the level of a group), or aggregate (operate for the whole population, but based upon analysis of many individual's data). Bowen and Filippini-Fantoni note how both personalization and customization may draw on explicitly generated user data (e.g., through questionnaires and registration forms), but that personalization also utilizes implicit information (e.g., through cookies and log files) (Bowen & Filippini-Fantoni, 2004). Thus, even if customization is seen as being largely manual, it may still be underpinned by digital platforms and personal data.

An alternative strategy is to encourage humans to differentiate experiences for each other, although again potentially with support from digital platforms. This more social approach to differentiation has been referred to as *interpersonalization*[1] by Eklund (2020), who first called attention to it as a result of her ethnographic studies of museum visiting which revealed the importance of social meaning making among visitors. Eklund calls for a shift from "designing personalized toward interpersonal experiences" and raises four design sensitivities that need to be considered: interpersonalized meaning-making, playful sociality, social information sharing, and social movement. While the approach could be seen as being as much about customization as personalization, the term would appear to capture the essential social dynamic of getting one person to differentiate an experience for another, and so we adopt it here in directly responding to her call.

Ekund's (and indeed, our own) interest in museums as a site for interpersonalization is not accidental. The modern museum faces many challenges including widening the audience demographic to include younger audiences, reaching out to those who have traditionally been excluded or have not seen museums as being relevant to their lives, and opening up to new voices and narratives as they struggle to deal with the legacy of colonialism. Many have been turning to digital technologies as a potential solution as they are perceived to be popular with younger demographics, associated with the wider world of social media, gaming and digital entertainment, and can be used to present multiple narratives around events. There has also been a growing awareness within the museum sector of the need to offer differentiated experiences to different audiences (Falk, 2009; Falk & Dierking, 2012) and the adoption of digital technologies brings with it the potential for achieving this through the above strategies of customization, personalization and now interpersonalization. A notable example of a customizable museum experience is the "Pen" device offered to visitors at the Cooper Hewitt, Smithsonian Design Museum, which allows visitors to digitally "collect" exhibits they encounter and engage in various co-creation and co-curation activities (Chan & Cope, 2015). The personalization of museum experiences has received widespread attention since the 1990s (Bowen & Filippini-Fantoni, 2004; Lynch, 2000; Oberlander et al., 1998; Paterno & Mancini, 1999; Stock et al., 1993) including projects that have explored how to design personalized experiences and exhibitions in museum contexts (Kontiza et al., 2018; Kuflik et al., 2011; Jonathan Lee & Paddon, 2017; van Tuijn & et al, 2016). As documented by Ardissono et al. (2012), early work on personalization in cultural heritage has focused largely on systems that adapt to individual users through user modelling aimed at matching users with relevant content. Continued interest in this area is demonstrated by the PATCH workshop series on Personalized Access to Cultural Heritage, currently in its 11th iteration. Recent research has continued to focus on issues such as user modelling and recommender systems (Almeshari et al., 2019; Castagnos et al., 2019; Dahroug et al., 2019; De Angelis et al., 2017; Deladiennee & Naudet, 2017; Fishwick, 2016; Katifori et al., 2019; Mauro, 2019; Mokatren et al., 2019; Sansonetti et al., 2019).

---

[1] Eklund's use of the term differs from previous uses in Psychology to refer to a shift in psychoanalytic theory that draws increased attention to the interpersonal relation between patient and therapist (Aron, 2001; Bonovitz, 2009) and in Education (e-learning) research to highlight the importance of interpersonal interaction and communication between people (Garrick et al., 2017, pp. 5–6; cf. also Oomen-Early et al., 2008), though the latter would appear to be somewhat related through its emphasis on the social.

However, the personalization of museum experiences is not always easy. Not and Petrelli (2019) suggest that a major obstacle to large-scale adoption of personalization in cultural heritage is the complexity of the technical systems, requiring technical expertise that is out of reach for most cultural heritage professionals with the implication that successful approaches may need to be technically lightweight. While personalization tends to be approached as a challenge of matching users with relevant content, it can also be considered as the challenge of *making an experience feel more personal*, or developing a personal connection between the visitor and the museum. Not et al. (2017) approach this challenge through a system for personalized text generation that creates personalized postcards summarizing the visit to the museum. Museums have also explored personalized storytelling (Katifori et al., 2014; Vayanou et al., 2014) and play (Vayanou et al., 2019). Marshall et al. (2015) report on an experiment in which museum objects were given personalities and made to "compete" for display based on which object could capture visitors' interest the most through both physical presence as well as interactions on Twitter. In the museum context it is also important to note that museum visits tend to be a social activity, and approaches to personalization therefore need to address the social context of the visit (Fosh et al., 2015, 2016, 2014; Lykourentzou et al., 2013; McManus, 1989).

In what follows, we explore a novel approach to the challenges of differentiating museum experiences, one that differs from both the conventional customization of websites and visits and also from automated personalization based on visitor profiles. We turn to the idea of interpersonalization and explore what traction this might offer for creating new kinds of visiting experience that meet the needs of the contemporary museum. We present two designs in which a museum visitor is tasked with creating an experience for another museum visitor, offering them the opportunity to see the exhibition through another person's eyes as it were (Spence et al., 2019). These designs offer concrete responses to the idea of interpersonalization. On the one hand, the adaptation is done manually by a human user, but on the other hand that person is adapting the experience for another person, carrying out a similar function to what a computational personalization system would do. However, the adaptation carried out by users in our two examples is qualitatively different from the personalization that can be done by a computer algorithm. A human user can employ their full range of intellectual, intuitive, emotional, social and expressive capabilities in order to create the most gratifying experience their imagination allows.

*Intimacy*

The second key theme that emerged from our paper is intimacy, which has also been considered by previous research in HCI. Generally, the word intimacy is associated with the private and emotional sphere of one's life. It is often used in relation to physical closeness or emotional investment in relationships, such as between romantic partners. However, more broadly, intimacy is used to describe a range of things happening at the local, micro level, as well as on embodied levels, and on levels that involve the psyche in one way or another (Wilson, 2016, p. 249). For instance, the term is occasionally used to describe experiences that take place in the encounter between people and their (living or nonliving) surroundings, such as referring to art or nature experiences. In the light of this, Sadowski suggests that in the broadest sense of the term, intimacy describes "a context that is relational, and that this relation affects one's body and embodied self" (Sadowski, 2016, p. 46). According to her,

"getting intimate with someone or something means crossing a boundary and connecting with the other, and being at risk of losing oneself to some degree" (Sadowski, 2016, p. 45).

Within HCI, it is sometimes acknowledged that the term is ambiguous, subjective and hard to define (e.g. Kaye et al., 2005). However, most often intimacy is used with reference to interpersonal relationships and research on interactive technologies to express, share and communicate already established intimate feelings (although there are exceptions such as the work done by Schiphorst et al. (2007)). The interest in computer-mediated intimacy goes back to before the turn of the millennium (Dodge, 1997). This and other early signs of interest in the topic are perhaps best represented in the 2003 Intimate Ubiquitous Computing workshop at Ubicomp (Bell et al., 2003). According to Gaver, technologies for mediating intimacy can be categorized into two groups: a) those which mediate intimate expressions and b) those which evoke intimate reactions (Gaver, 2002). In the first case, technologies are used to reproduce intimate action or situation (Counts & Fellheimer, 2004; Goodman & Misilim, 2003; Markopoulos et al., 2004; Mueller et al., 2005) and in the second, the technologies rely on materials and abstract representations as a way to elicit feelings of intimacy between family members, romantic partners, friends or even complete strangers (Chang et al., 2001; Gaver & Strong, 1996; Schiphorst et al., 2007; Tollmar et al., 2000). When it comes to design strategies, approaches that utilize the expressive, evocative and poetic capacities of electronic media are often argued for (Gaver & Strong, 1996; Grivas, 2006). For instance, Jayne Wallace has used digital art and jewelry to explore issues such as experiences of enchantment through the evocation of intimate rituals (McCarthy et al., 2006), how aesthetic experiences including the digital connection to another place can lead to feelings of interpersonal closeness and intimacy (Wright et al., 2008), and how a sense of self, home and intimacy can be enabled for people living with dementia (Wallace et al.,

2012). These design explorations emphasize a pragmatist aesthetics of interaction (Wright et al., 2008) wherein the intellectual, sensual, and emotional are equally embraced. Moreover, it puts the relational and dialogical aspects of experience into focus, acknowledging how self, object, and setting are actively constructed and how the dialogue between them plays an important role in completing any form of designed experience.

However, contrary to the aesthetically rich approaches employed by Wallace and others, it has also proven effective to build on the culturally and socially embedded nature of communication even in the case of extremely minimalist design. A study of "Minimal Intimate Objects", low-bandwidth devices for communicating intimacy for couples in long-distance relationships, revealed that "a single bit of communication can leverage an enormous amount of social, cultural and emotional capital, giving it a significance far greater than its bandwidth would seem to suggest" (Kaye, 2006, p. 367). Here, the constrained nature of the communication provided space for complex and evocative interpretations based on the partners' shared understandings of each other. Thus, the experience of intimacy relied on the richness of the relationship, rather than content or the visual appeal of the design.

Another form of intimacy to be found in the HCI literature concerns "vicarious" experiences, specifically applying digital technologies to give one person a close-up and intimate view of another's experience. A notable example involved the riders of amusement rides wearing a head-mounted video camera and microphone as well as heart rate and sweat sensors, with the captured data being broadcast to watching spectators who could tune in to an unusually close view of someone having an thrilling experience (Schnädelbach et al., 2008). The designer followed this up with a more intimate pairwise experience in which some family members watched from a distance as others explored a "horror maze" at a major theme park. These experiences proved to be intense and emotional, even at times challenging and uncomfortable. Indeed, creating an acceptable level of temporary discomfort was one of the key design strategies employed by the designer, an example of the more general concept of "uncomfortable interactions" in which intimacy, along with lack of control and visceral and cultural discomfort, is designed into experiences so as to make them entertaining, enlightening or socially bonding (Benford et al., 2012). Vicarious experiences have also been explored directly within the museum context, as for example in the Sotto Voce tour guide that enabled visitors to eavesdrop on other visitors' tours (Aoki et al., 2002).

However, associations between intimacy and discomfort are not restricted to overtly scary or thrilling experiences, but can also be found in more everyday situations. Fosh et al's (2014) study of museum gifting involved the receiver experiencing their gifts in the presence of the giver, which led to reported moments of awkwardness and embarrassment when gifts had not been well judged, appeared to convey inappropriate sentiments, and/or were not acknowledged appropriately. A study of The Rough Mile, a locative experience designed to give and receive music tracks, included a broadly similar account of a gift backfiring (Spence et al., 2017). Indeed, everyday gifting is a socially important and complex phenomena that consequently comes loaded with risks for losing face, both for the giver due to an ill-judged gift and the receiver arising from an ill-judged response (Sherry et al., 1993; Sunwolf, 2006). In short, while intimacy can deliver powerful experiences, it comes along with the risks of also creating awkward ones and so needs to be treated with caution as we explored in our two designs.

**Methodology**

The two case studies that we consider in this paper were two separate sub-projects undertaken in parallel within an overarching three-year long multi-partner European research project called GIFT (Back et al., 2018; Løvlie et al., 2019; Waern & Løvlie, in press). Both projects followed the approach of Research through Design, a design- and practice-led approach in which research findings emerge from reflections on the practical activities of designing and making. Reflection may involve critically appraising a portfolio of similar designs (Bowers, 2012; Gaver, 2012) to draw out common themes. Our approach also involves a significant element of Research in the Wild (Rogers & Marshall, 2017), as we examine two designs that were deployed in actual museums with public audiences under realistic conditions and studied what unfolded. Both designs also incorporated performative elements, with *Never let me go* harnessing the idea of one visitor controlling another's somewhat performative interactions in the public space of a museum, while the *Gift* app was designed by professional artists with a background in performance who were interested in bringing their artistic sensibility to the design of a mainstream visiting experience through the design of performative instructions as we discuss below. Consequently, our approach also incorporated elements of "performance-led research in the wild" (Benford et al., 2013), even though the two experiences were not overtly framed as performances.

Our reflections therefore encompass both the designers' and users' (visitors') perspectives, reaching out beyond the "design studio" to also consider the experience of real-world deployments in museums. The data-capture element of the *Gift* app relied on documentation of iterative designs, design meetings, and in-depth interviews with visitors during an initial prototype deployment at the Brighton Museum and Art Gallery in the UK in

July 2018. 57 users completed an exit questionnaire and a further 57 undertook a full interview. Researchers also gathered data from usage of the final version released in 2019, which is available through https://gifting.digital/gift-experience/, as well as a small number of in-depth interviews. These indicate that the final version has not significantly altered the reactions received in 2018. All user names have been pseudonymized.

*Never let me go* was part of a PhD research project that involved designing and deploying several prototype experiences. Here insights from the design process were combined with data gathered from four prototype deployments. Early iterations of the app were tested at three different art museums in Copenhagen with 6 users in total. The main trial was conducted at the National Gallery of Denmark. In total, 20 people of 13 different nationalities (mostly European) took part in the trial. 6 out of the 10 pairs were romantic couples; 1 pair were siblings; 2 were friends and 1 pair had just met for the first time. During the trial the participants were observed and photographed (with consent given beforehand) by a researcher, and afterwards in-depth interviews were carried out with them in pairs. The interviews (each between 30-40 minutes long) were recorded, transcribed and analysed. The observing researcher took notes continuously of what the participants were doing and at what time. Photographs were taken to supplement the field notes and to contribute to the overall impression of the trial.

As our two case studies were part of a common research project, reflections and comparisons between them occurred informally throughout their development, with the wider project team regularly gathering together to share and compare results with a view to developing common guidelines, tools and platforms. However, findings from both were initially published independently. The design and study of the Gift app was reported in a paper at ACM CHI 2019 (Spence et al., 2019) that focused on the idea of how visitors came to see the museum's collection through "others' eyes" (see below). Although some elements of our findings in this article dovetail with our previously reported findings, we use this article to deepen and extend discussions of interpersonalization, intimacy, the role of museums in app usage, as well as opportunities and challenges with realising such design strategies. The design and study of *Never let me go* was first reported in a paper for ACM CHI 2020 (Ryding, 2020), which focused on the design's combination of introspection and social play. Later, a paper with a more in-depth take on the relational aspects of the play design was published in ACM DIS 2020 (Ryding & Fritsch, 2020).

In this current article, we report for the first time new reflections across the two case studies that for the most part followed the end of the project, after there had been time to reappraise the work. This also involved revisiting the data captured from earlier studies and re-analysing it in the light of new themes that had emerged over the course of a series of discussions and collaborative writing.

We now briefly introduce our two case studies in order to help orient the reader to the unusual kinds of visitor experience that we are considering here, before then introducing our three main themes: Interpersonalization, intimacy and interpretation.

**The *Gift* web app**

The *Gift* web app is a primarily voice-driven, artistically crafted experience that enables museum visitors to create, give, and receive digital gifts from within the museum collections (see Figure 1, and supplementary video at https://vimeo.com/298647523/8679ad1d99). Visitors access the web app on a smartphone or a tablet, preferably using headphones. When visitors enter the web app they are greeted by a female voice speaking gently in an intimate tone of voice and a style more evocative of a personal conversation between friends than anything one might expect to hear from a public institution:

> Today you're going to make a gift for someone special. They might be next to you right now. They might be on the other side of the world. Close your eyes and try to get a picture of them in your head.

The narration frames the experience in terms of thoughtfulness and care (Figure 2). The giver is prompted to take photos (where permitted) of the objects they wish to include in their gift, and record an audio message. If they feel uncomfortable speaking in front of the object, they can find a more discreet place (Figure 3). They may then repeat this process, if they wish, for a second and third object to be included in the gift. Once completed, the giver can send the gift to the receiver via a link embedded into a messaging service such as SMS, email, WhatsApp or Messenger.

| Giver | Receiver |
|---|---|
| 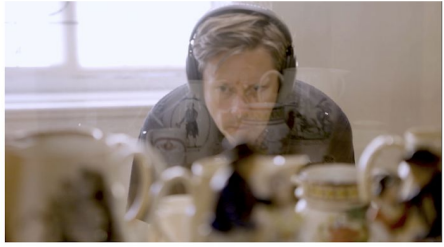 | 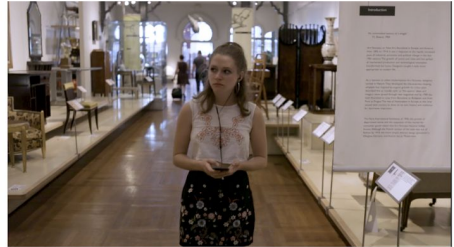 |
| Think of someone and search the museum for items they would like | Search out the items you have been given |
| 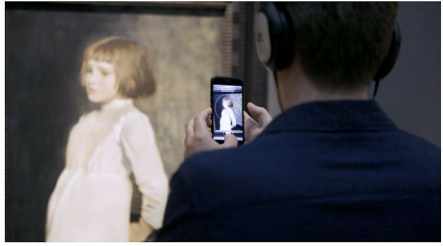 | 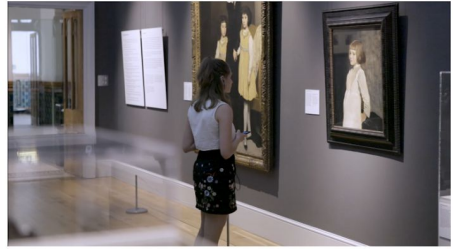 |
| Take photos so they can find them | Listen to the messages |
| 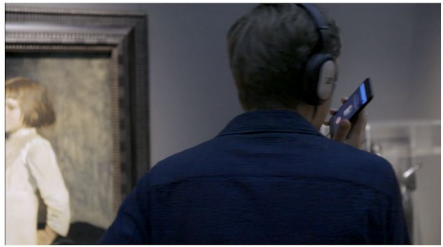 | 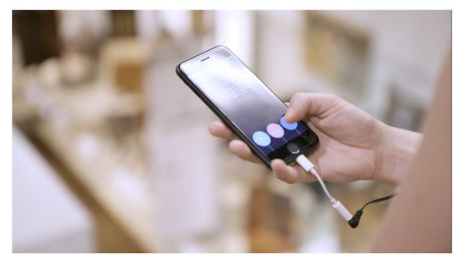 |
| Record personal messages | Record your response |

Figure 1. Overview of the *Gift* app experience. Photos by Charlie Johnson.

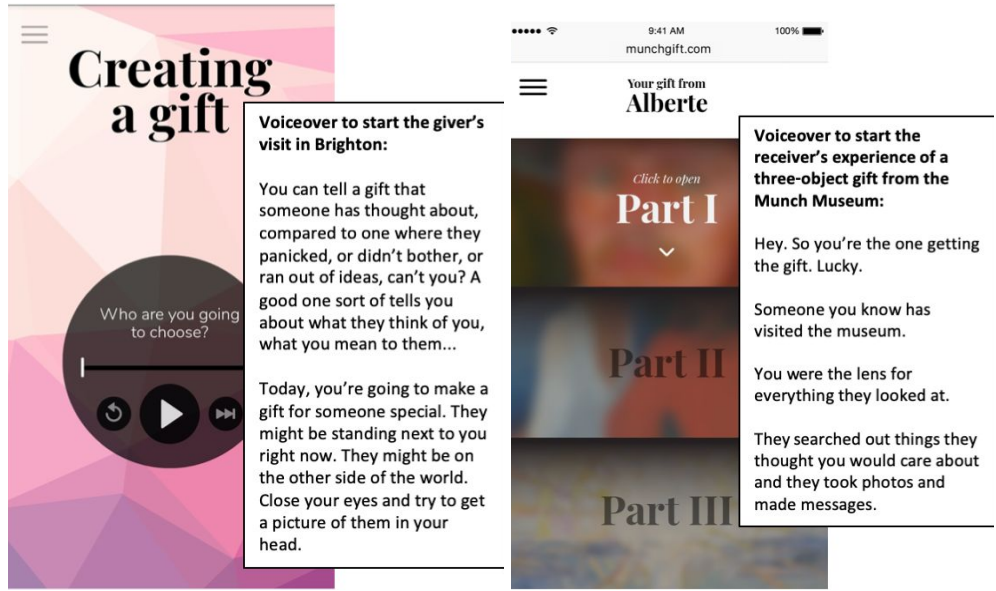

Figure 2. The beginnings of the giver and receiver experiences annotated with corresponding voiceovers. (This figure shows the interface and narration in the final versions released in 2019.)

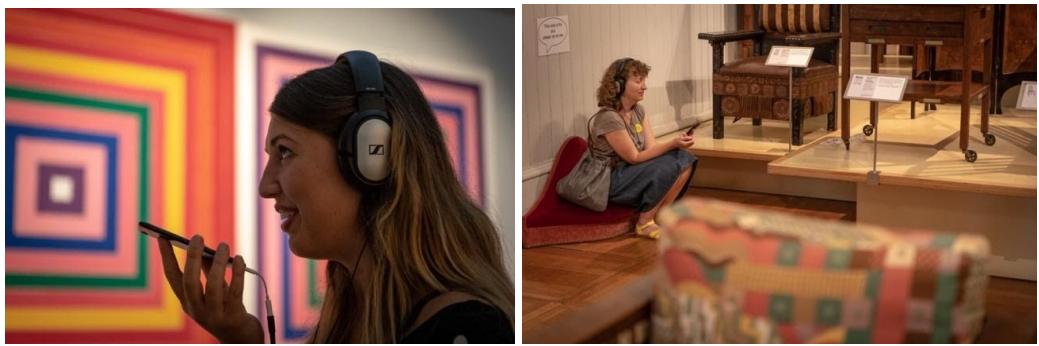

Figure 3. Speaking and listening in the museum space. Photos by Charlie Johnson.

When receivers click the link to their gift, they are taken to the unique web page for their gift and hear the same narrator's voice orienting them to this unusual gift-receiving experience. If they receive their gift at the museum where it was made, they can follow clues input by the giver to locate the objects in the gift and listen to the recording while standing in front of the objects. If they cannot attend the museum in person they can still read the clue, see the photograph, and hear the message. In turn, they are asked to record a response for their giver. These gifts can feel like a personalized museum tour, a museum-style "mixtape", a collection of hybrid objects, or something else entirely, depending on the receiver's individual perspective. Text entry and visual interactions are kept to a minimum so that visitors can keep their visual attention on the museum and its objects, while their mental and/or emotional attention is directed at their friend.

The web app was created by Blast Theory based on foundational research by Lesley Fosh and colleagues into the approach of gifting personal interpretations, first between couples in art galleries (Fosh et al., 2014) and then among small groups of families and friends in museums (Fosh et al., 2016). These exploratory studies employed low-fi prototyping and observation to reveal how gifting might help tackle two key challenges faced by museum curators: encouraging visitors to make their own interpretations of the objects they encounter, and personalizing the visiting experience. Blast Theory extended this theoretical approach through a collaboration with researchers at the University of Nottingham and IT University of Copenhagen (including the authors) over the course of three years' experimentation with concepts, combinations of user groups, and modes of gift composition, resulting in the web-based app described above. The web app was created in collaboration with Brighton Museum and has later been commissioned by the Munch Museum in Oslo, Norway, and the Museum of Applied Art in Belgrade, Serbia.

Our study of the first public deployment (Spence et al., 2019) revealed how the experience led many visitors to see the museum "through other eyes", either the giver through the receiver's eyes or the receiver through the giver's eyes. The experience as a whole tended to feel very different from a traditional audio guide, more often than not making an emotional impact of some sort based on the sense of connection cultivated by the app's design. In terms of design aims, Blast Theory considered both personalization and interpretation of the museum experience to hinge on the primacy of visitor-visitor relationships, folding the visitor's experience and interpretation of museum objects into the larger aim of supporting the visitor-visitor relationship. And although a visitor could choose anyone to receive their gift, they were prompted from the outset to think in terms of "someone special", someone for whom they would want to invest time and effort. All elements of the app were designed to support that premise.

Blast Theory's Lead Artist for the project in 2018, John Hunter, describes their aim for the voice narration in this way: "It allows her to put herself in the same boat with you and create a sort of instant familiarity" (Hunter, 2018, author interview). The user's spoken elements, in turn, formed part of the design element that they intended to make each visitor's gift "something that could feel meaningfully personalized rather than arbitrarily personalized" (Hunter, 2018, author interview). This can also be phrased as Blast Theory's adopting a strategy of "recipient design" (Fosh et al., 2014, p. 632), meaning that the selection and interpretation of exhibits was oriented towards a specific recipient rather than towards an official interpretation from the museum or a more general visitor demographic or persona, as is often the case with contemporary museums and cultural heritage institutions (Goulding, 2000). From the outset, the app was designed to scaffold, not dictate, the giver's interaction with the museum's contents in order to invite interpersonal interpretations that would have strong emotional, embodied, and experiential characteristics for each individual involved, and that would possibly reflect and possibly impact their relationship (for support in the gifting literature, see e.g. Camerer, 1988; Lawler & Yoon, 1993; Richins, 1994; Ruth et al., 1999; Sherry, 1983).

The feedback gained from the deployment in 2018 led the 2019 iteration to explore opportunities for meaningful personalization still further. Blast Theory considered the value of a fully in-the-wild digital proposition at a museum, what value a visitor would derive from the experience of using the app, and what value a gift receiver would derive from the gift. They took their cue from Kevin Bacon, Digital Manager of the Brighton Museum and Art Gallery, with whom they had developed a close working relationship over the three years of iteration. "One of Kevin's observations about digital experience in museums is that it's not around trying to sort out more content or more activities to museums, it's finding a way of focusing people's attention so they're not overwhelmed by the amount of content that's already there," said Nick Tandavanitj, one of Blast Theory's lead artists (2019, author interview). In turn, they focused the app even more tightly on the idea of presenting a handful of objects, or even just one, but to use every means at their disposal to let gift-givers put their own unique mark on it that might change the way that their receiver saw it or felt about it – and might change the giver's own experience at the same time. They did this partly by streamlining the user interface and the narrator's text, but also partly by including a limited selection of sharing mechanisms embedded in the gift-giver's or gift-receiver's own device.

> When we chose social channels, we explicitly chose ones that were private messaging channels as opposed to publishing channels because everything about the setup for it is to say this is for an individual, and about reflecting on that single person. … That's the value those channels give to the messages that you receive (Tandavanitj, 2019, author interview).

**Never Let Me Go**

*Never let me go* is a two-player experience, in which visitors can playfully guide a companion through the museum. It provides two roles: The Avatar and the Controller. The Controller is given the tools to influence or shape the Avatar's experience, as both players explore the exhibitions together (Figure 4). The system consists of two interconnected web apps. Whilst the Avatar never really interacts with the app (except to press START), the Controller uses the app to send different commands, questions or instructions to the Avatar, who receives them as pre-recorded voice messages. All audio is played for both players simultaneously, in order for Controllers to get a sense of what the Avatar is experiencing.

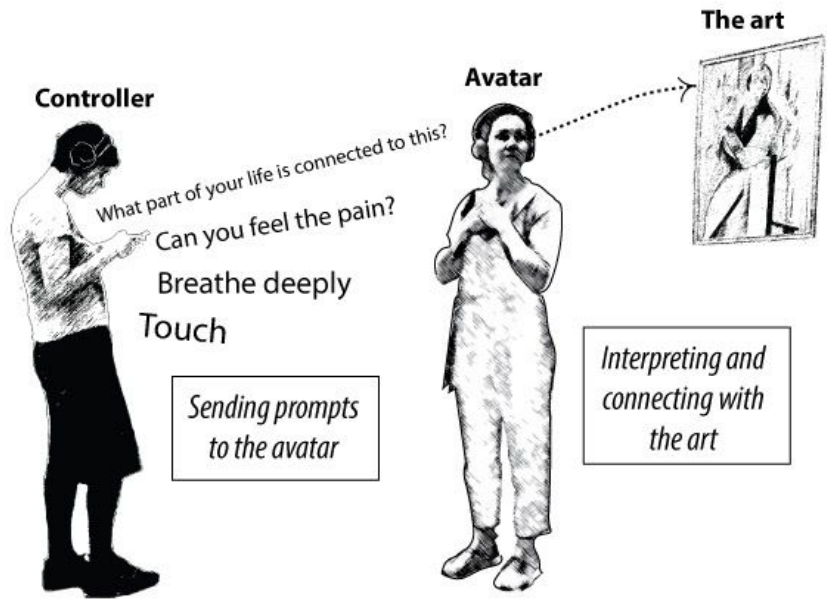

Figure 4. An illustration of how *Never let me go* works.

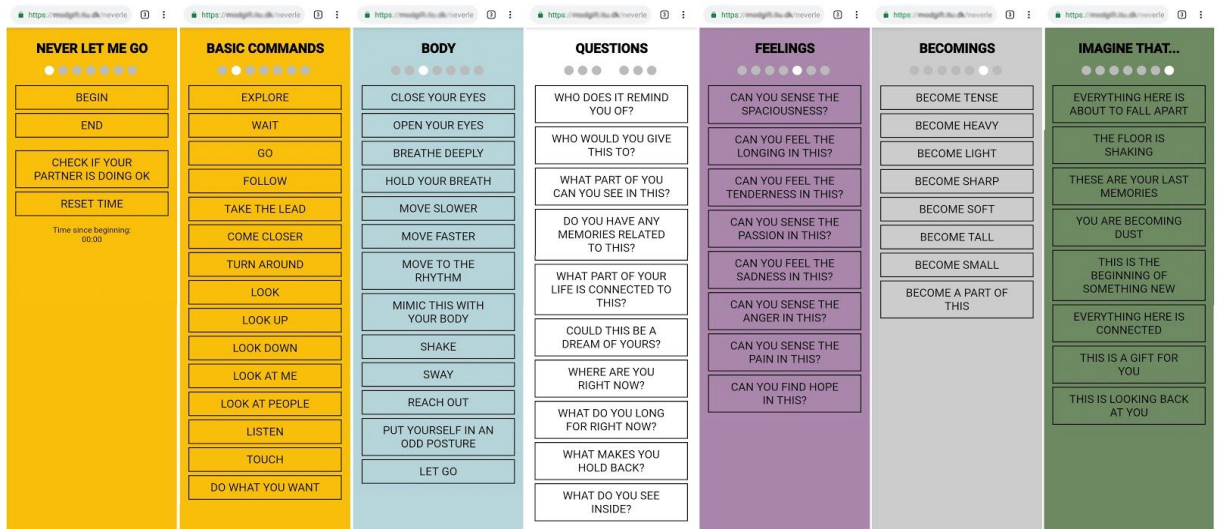

Figure 5. Screenshots from the Controller app.

In the Controller app, there are six different categories of prompts to choose from. The first category called "Basic commands" consists of direct prompts such as "Explore", "Go", and "Turn around". The purpose was to facilitate movement and exploration of the museums and its exhibitions. The second is called "Body" and consists of instructions relating to the body of the Avatar, such as "Close your eyes", "Hold your breath" or "Mimic this with your body". This was included to encourage the participants to have a more embodied approach to the museum experience. The third category consists of personal questions that could be used in relation to the art, for example "What does it remind you of?" and "Who would you give this to?" The idea behind this category was to encourage introspection and emotional connections with the artwork. The fourth category is "Feelings" which consists of questions again to be related to the artworks, but this time in order to direct the Avatar's attention to the emotional content of an art piece. Examples are "Can you feel the tenderness in this?" or "Can you sense the anger in this?" The fifth category is called "Becomings" and consists of prompts that are deliberately ambiguous and open for interpretation. Examples are "Become heavy", "Become small" and "Become part of this." As with the "Body" category, these prompts were included for participants to explore new ways of being in the museum. Lastly, there is a category called "Imagine that". This consists of instructions intended to trigger the Avatar's imagination. The idea was both to facilitate narrative play and to induce a sense of urgency in order to intensify the Avatar's experience. Examples of this category are "Imagine that these are your last memories" and "Imagine that everything here is connected." Apart from the categories described, there is also a "Begin" and an "End" option in the menu. These trigger voice recordings with the purpose to frame the experience and give the players an idea of what to expect from each other.

The design of *Never let me go* was inspired by Blast theory and their work with the *Gift* app. In a similar manner, it was motivated by the task to design a (more or less) generic mobile app which could be used in any large to mid-size art museum, gallery or sculpture park. However, it was developed by Karin Ryding as part of her PhD project with a research agenda that involved exploring how museum experiences can be affectively enhanced through play. Therefore, whilst making use of some of the same components as the *Gift* app (such as using voice to give instructions), a ludic design approach was additionally employed. As a result, the design of *Never let me go* put a stronger emphasis on performative and corporeal qualities, as well as the relational dynamics which become naturally emergent as both players are physically present in the museum. The design strategy included the use of a certain level of ambiguity, in both content and player roles, in order to give room for curiosity, shared exploration and play. In a previously published paper, the notion of using play design as a *relational* strategy to intensify affective encounters in art museums has been put forward (Ryding & Fritsch, 2020). To a certain extent, this article works as an extension or broadening of that same discussion by including not only social play but also the perspectives of interpersonalization and intimacy.

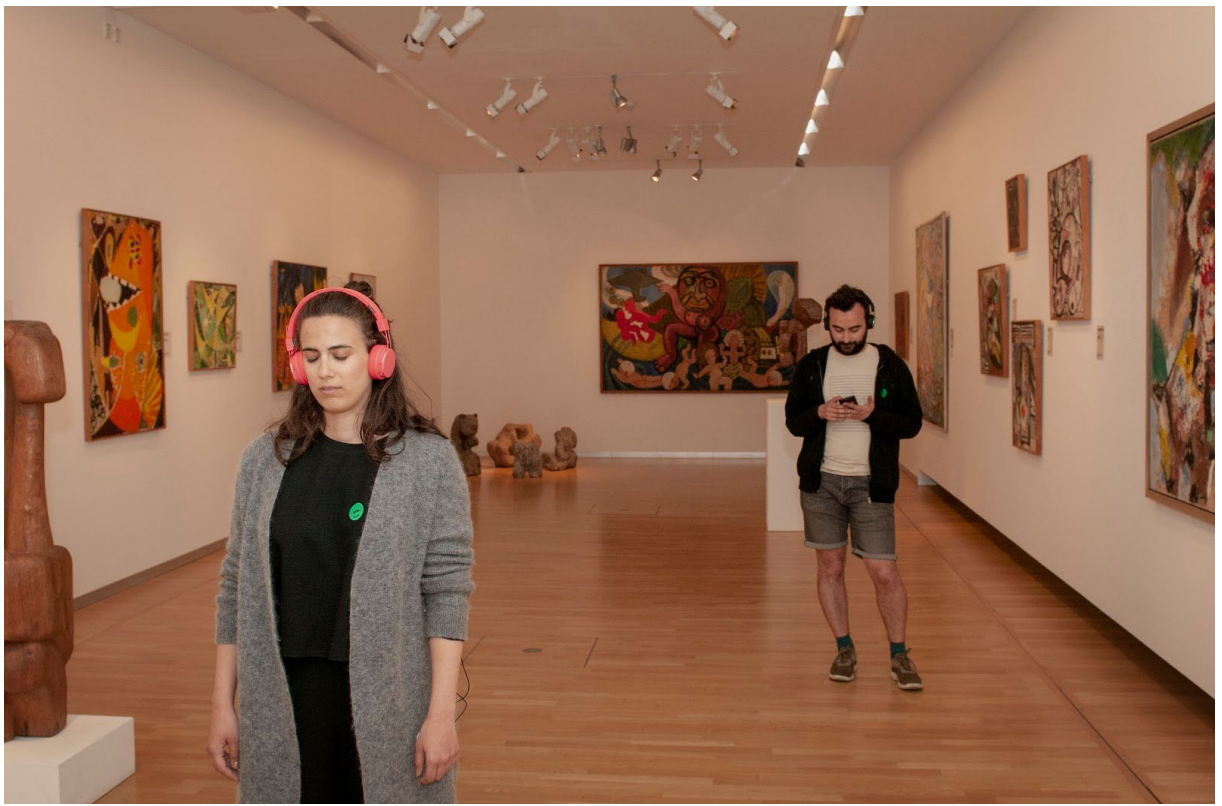

Figure 6. An Avatar being prompted by the Controller to close her eyes. Photo by Johan Peter Jønsson.

The main trial of *Never let me go* took place between April 22 and May 2, 2019, at the National Gallery of Denmark. The participants were recruited beforehand through public invitations on social media, and from a mailing list for people interested in cultural experiences in the Copenhagen area. The findings, which are presented in more detail in (Ryding, 2020) and (Ryding & Fritsch, 2020), show that participants found it to be a highly immersive experience, which was more personal, emotional and sensuous than a regular museum visit. Playful moments of teasing and laughing became naturally intertwined with more serious moments of introspection. One of the interesting advantages of the impromptu approach provided by the design, was the possibility for players to use whatever was happening in the present moment. When it worked at its very best, this led to emotional experiences where the conditions seemed to fall almost perfectly into place in a surprising, almost magical, way – close to what is called serendipity (Makri & Blandford, 2011). However, there were also moments when players felt disconnected or distracted from the exhibitions as their attention was drawn towards each other.

**Interpersonalization**

In both *Gift* and *Never let me go,* visitors are tasked with crafting an experience for another person – either the receiver of the gift, or the Avatar. In so doing, they may be said to be personalizing the experience for the other person – using their knowledge of the other person's interests and preferences to create and facilitate an experience they think the other may enjoy.

In *Gift,* the gift-givers are explicitly tasked with creating a personalized experience of the museum for the receiver: selecting objects based on what they believe the other person would like, and subsequently presenting them for the receiver with a photo and a personal audio message which addresses both the object and why this object is relevant as a gift for that receiver and that receiver alone. Some participants, including receivers, spoke explicitly in terms of personalization. For example, Helen, who both gave and received, commented without prompting that "it's nice how you can personalize it for different relationships, what they'd like around [the] museum." Interviews with participants revealed that gift-givers spent considerable effort searching for the right objects to give. However, both givers and receivers tended to value the gift more for the meaning it expressed about their connection with each other through the selection and justification of the gift's objects than for the objects *per se*. George described the type of experience that many givers reported:

> Two of the items I chose were paintings which linked me and my daughter together and our own histories. I would probably have done this anyway – paintings of children always make me think of my own children – but it was more special as I knew that she would see them too.

In *Never let me go* the personalization took place on several levels at once, leading to experiences that could be described as affectively engaging. First of all, as Avatars, the players would be guided through the museum by someone they knew and trusted, which opened up for new possibilities in terms of more personal and playful connections with the exhibitions and the architecture. The Controllers, on the other hand, focused on making meaningful experiences for their Avatars, but by doing so they also connected with the museum in a personal way. In the words of Anna:

> I wanted to see it as a way to share, like a feeling or a situation, wordlessly. You keep it separate and private, but you could still express: "This is something I enjoy. I like to think about spaciousness here. And now I make you think about it too, and hopefully you will enjoy it as well."

Test participants put effort into crafting the best possible experience for the receiver/Avatar:

> My only concern was to build the prompts in doing something that would be a cohesive and interesting experience. I was not concerned about giving too much orders but just about having a sense of progression or having something interesting. Not just random things or I'm going to make you do stupid things just because I can. (Peter)

The personalization that is done in these two designs is qualitatively different both from systems-driven personalization and user-driven customization, as discussed under related work. Rather, it reflects Eklund's concept of interpersonalization. An important difference between this and customization is that the object that is being adapted – the museum exhibition – is being appropriated by the visitor to be used in interpersonal communication and play. Thus, the experience of the exhibition becomes intertwined with the interaction between the visitors. In *Gift*, the givers are not just customizing an object but crafting an experience and initiating a dialogue with the receiver. In *Never let me go,* the foregrounding of the personal relation becomes even more evident, as it is entirely up to the players when and how to engage with the museum exhibition.

Thus the personal aspects of *Gift* and *Never let me go* do not flow in one direction from giver to receiver or from Controller to Avatar, but instead form a process of interpersonalization that affects both participants in the exchange. The experience reflects both the giver and receiver and is personalized for and by both: *"It makes you experience the museum in a different way. It's a new point of view. You have fun, similar to the fun when you play a game or share time with somebody" (*Gift* app user Lavender). The *Gift* app invokes elements of the complex social ritual of gift-giving, which comes with expectations and gratifications for both the giver and the receiver. The giver must exert appropriate effort to create a good gift, and make sure that this effort is evident in the gift. Ideally the gift should reflect the relation between the giver and receiver, whether it manifests itself in a deeply meaningful and touching moment of closeness, a simple joke, both, or anywhere in between. In return, the giver may receive gratitude and, hopefully, experience a strengthened bond with the receiver. The experience for both giver and receiver is mutually shaped by the relation between them, as in a dialogue. Similarly, in *Never let me go*, the Controller is handed the power – and the responsibility – to shape the experience for the Avatar. Interviews with test players demonstrate that many felt a strong sense of responsibility/expectation: "You get the feeling of having a responsibility and you feel like you want the experience to be good or interesting for the other person" (Laura). The receiver and Avatar contribute to the interaction by their reactions to the gifts or commands they receive and how they choose to respond to them. Following the commands allows the Avatars to explore and play with the boundaries for behaviour in museums: "At times of course getting instruction gives you an alibi. But especially with the physical prompts I was limiting myself to what I feel is acceptable behaviour. Without any onlookers I might have done stuff bigger" (Lisa). While the system does not offer the Avatar any feedback mechanism through the mobile interface, the Avatars explored different ways to relate back to the Controller – from relying on discreet smiles, taking off the headphones and talking back, or playfully misinterpreting commands:

> You came next to me and said: "Come closer." I knew, I was sure that you meant to go closer to the painting, but I thought I'm not going to go closer to the painting. I'm going to go closer to her and make her uncomfortable. That was fun. (Cetin)

One interesting aspect of the interpersonalization offered by *Gift* and *Never let me go* is that they offer visitors a choice of roles: Giver or receiver, Controller or Avatar. In many cases participants have taken turns to try out both roles – e.g. they may give a gift to someone they came to the museum with, and receive one in return. If no one had already made a gift specifically for the visitor, they might instead choose to receive a generic gift created by museum curators. Although curators could not build on personal relationships with unknown visitors, the gifts were made with emotionally driven choices, personal explanations, and a similar style of delivery as many personal gifts. Within the overall role of "giver" we have noted several interpretations of that role. Some users created several gifts for others without receiving any in return, whereas some made only one. Those who arrived in pairs or small groups usually gave to each other and only occasionally gave an additional gift to someone not present. Many were excited to imagine their receiver getting the gift, while a few even created gifts for themselves. In the *Never let me go* test sessions, the participants were instructed to switch roles. However, in post-experience interviews some participants expressed a preference for one or the other role, indicating that if they were free to choose they might be more likely to take that role. This seemed to be based on a combination between their personality and the mood they were in. As Alex explains in relation to the Avatar role, "I realized that I'm very strong willed. So, I just want to go where I want to go and look. And now it was like oh I have to relate to what somebody is telling me to do". Whilst Lisa says, "I liked somebody else being in control. I'm in control of a lot of things when I'm at work and I was a little bit stressed before I came here. So, this was really nice."

Users of *Gift* and *Never let me go* employed a variety of strategies for creating satisfactory experiences for their counterparts. For instance, creators of gifts might employ different rhetorical styles of communication, ranging from playful and jokey to contemplative and reflective. Similarly, *Never let me go* players could use a variety of "play styles", such as playing with norms and boundaries (e.g. instructing Avatars to imitate the artworks with their bodies), trying to set up interesting or comic situations, or inviting deep contemplation. The instructions and commands in both apps were sufficiently open to interpretation to facilitate a spectrum of styles of engagement. As Peter explains:

> In terms of the playing element it was very dependent on the art. When the art became let's say very modern to a point where I couldn't connect with it. The playfulness became a defence mechanism. As I don't understand this, I will make fun with it. Because if I can't really connect with it or interact with it on an emotional level. Then I can at least make a fun experience out of it.

It is worth noting that the continuous oscillation between being serious and playful seemed perfectly natural to the players, reflecting their already established relationship dynamics.

The interpersonalization described here has some clear limitations in comparison to algorithmic personalization: in particular, it requires two users to collaborate, and relies on users investing significant time and effort. However, this approach also holds some advantages. First, as it relies on the relationship between two people, it allows for deep and meaningful connections that draw on the power of the interpersonal relationship. Second, as described above, the experience can affect both the people involved. As such, interpersonalization may benefit two people for each interaction. Third, this approach can be implemented without collecting personal data about the users, which may become an important advantage, as many users are increasingly concerned about privacy and protection of personal data.

**Intimacy**

We now consider how the kinds of interpersonalization described above and embodied in our two case studies critically depends on establishing a degree of intimacy between pairs of visitors, which in turn leads to a deeper sense of intimacy with the museum itself.

*Intimacy between visitors*

Our two case studies established different kinds of interpersonal relationships between visitors, both of which involved a degree of intimacy, albeit in different ways.

Gifting can be an intimate social experience, often conducted between close friends, family members and romantic partners, and may serve to reflect or strengthen social bonds. It may also rely, at least when done well, on a relatively intimate knowledge of the other in terms of the kinds of gifts that they would appreciate. Our study of the *Gift* app revealed how givers often create gifts for those whom they feel very close to, such as romantic partners, siblings, parents, children, or best friends. They then choose objects based on personal knowledge of their receiver and leverage the intimacy of their relationship to create personally meaningful connections between receiver and object, or in a three-way relationship between giver, receiver, and artwork. There were numerous examples of giving intimate gifts, typified by the following selections. Teenager Jack, on giving a gift to his girlfriend Lolly, reported, "I thought about what they liked, what their personal tastes were and also what would make them laugh if they found it as a gift." Lolly, on receiving Jack's gift, observed, "It was really thoughtful of him...like, taking inspiration for something he's trying to make for me, something he knows I've wanted for a long time, showing me, like, an example, something if he could create it he would for me." Gordon made a gift for his grandson: "My grandson has sent me his first painting he'd done, which I had gotten over email, and I just thought I'd send him something to do with art..." Teenager Kristin, making a gift for her mother, said: "Some things stood out more. I thought that's exactly what my mum would like, and so reading information about it made it feel more close to me or to my mum. So I connected my mum with this item and I think that helps remembering information better."

We also see intimacy in responses from receivers. For example, Helen described her feelings about receiving her gift (given in exchange with her sister, who took part at the same time) as "excited when I open – when I found my gift. But I felt, I felt happier when I heard [the giver's] voice, really." Receivers tended not to want to share their gifts, either. Those few who did express an interest in sharing them would do so only with other intimate friends or family members. As Adam put it, "I feel like mine was very specific. I couldn't send it to someone else [because it related to something in our shared past]." The intimacy inspired by *Gift* can occasionally be conveyed by the image as much as the voice: "It was really really lovely and quite touching, and it was a good conversation-starter. And it does say something about what the other person thinks of you in what they choose, and they don't necessarily say in words." This adult participant, Carol, had received a gift from her father, a man who did not seem chatty or overtly affectionate towards his daughter. Asked how she felt about their relationship after using *Gift*, she replied, "Closer," and described it as "a bonding experience." These levels of intimacy were by no means universal: many described the experience as simply "fun" (Pat), "just sharing something" (Gill), "confusing" (Wayne), or even "restrictive" (Susan). However, intimacy of some sort was one of the most common terms of reference for the app, and people who had strongly positive experiences spoke almost exclusively in terms of intimacy or similar relational feelings.

In *Never let me go* the setup with two roles, the Avatar and the Controller, provided a specific form of power relation between players, but one that also appeared to strengthen feelings of intimacy between the two. As the Avatar, they would put themselves into the hands of the Controller, making themselves vulnerable to a certain degree. Controllers, on the other hand, would accept the challenge of being the person in charge of the situation which also entailed being exposed to critique. These shifts in agency led to a special bond being created between the two players, as they explored the museum together. As Alex puts it: "If it's with somebody that you know well, it gives a certain framework and certain ways to exchange." The prompts also provided the opportunity for players to play with intimacy in the form of induced introspection. For example, by using the available questions in relation to specific artworks, Controllers were able to trigger very personal moments for their Avatars. Lisa gives this example:

> We were looking at a painting similar to "The Last Supper." First, I asked: "Imagine that this is looking back at you." And then I followed up with "What does it remind you of?" Because nobody in the painting looks at you, they're all kind of looking sideways, my partner had this experience that he was being isolated because nobody was looking at him. That brought up some personal memories from his youth. So that was really unexpected and a personal moment and revealing some reflections.

Sharing intimate moments, in this way, was a matter of trust. It was the couple's individual relationship that set the frame for how the intimacy was perceived. If the two players were already in an intimate relationship, this might strengthen the experiences in many ways. As Laura says, "I think it is easier when you know the person well. So, you have an idea that okay this is going to make them react." On the other hand, if the couple hadn't established that level of intimacy beforehand, it could be interpreted as inappropriate to ask this type of personal question. The *Gift* app's giver-receiver pairs did not place themselves in each other's control quite so overtly and could often receive their gifts on their own, so issues of trust emerged rarely and then only in a positive sense, as in Josephine's description of her feelings towards her giver after using *Gift*: "I still feel the same way in the sense that I'm in love with him. I trust him and I feel understood, I guess." Two others said that the objects they chose spoke to or reflected the trust they feel in their relationship with their receiver: Katalin described how "I sent a picture to my friend. To me, it means a close friendship, also trust, thankfulness"; and Lindsay's objects spoke to the "good sense of humour" and "deep trust there." Finally, Mark felt trust in the app's "instruction" to go with their instincts and wander the museum until something called out to them. "I just trusted, yes" – and that trust paid off in a worthwhile gifting and visiting experience.

Trust is important to helping negotiate various risks that come along with the use of intimacy. One is of experiences being uncomfortably intimate. As Lilly of *Never let me go* commented: "Some of the questions were too intimate. I felt that those questions were leading more toward deeper feelings and memories. Like when you ask them in that way, in an art setting. I don't know. It felt weird." The idea that other museum visitors could be watching was also a source of discomfort. As Michael of *Never let me go* says:

> It felt somewhat awkward, because I was aware of people around the room. Probably they were not looking at all or they were minding their own business. But it's part of most people's common thoughts, no? How am I being perceived or am I acting out of place.

Quite a few visitors to Brighton Museum found it uncomfortable to speak into their phones inside the museum, at least at first, and sought not to draw the attention of other visitors or museum staff to themselves. Many of these participants found their recordings of second and third gifted objects to pose far less discomfort. The intimacy of the experience was able to overcome this obstacle for one receiver in particular in a remarkable way. Natasha found the experience both "moving" and "touching." For her, "somehow the phone was, it was conveying something precious to me. So… having had a discomfort with it, initially, it went away and turned, and became a positive, a very positive thing". We received no feedback on inappropriate or otherwise unacceptable gifts that impacted negatively on personal relationships, though to be fair, this may be because the short and location-bound nature of our 2018 deployment allowed us far more access to givers than to receivers.

*Intimacy with the museum*

While our two experiences evidently hinge on intimacy between visitors, a second and different sense of intimacy also emerged from our studies in the form of greater intimacy with the museum itself. Rather than encouraging visitors to undertake wide ranging explorations of the entire museum or to try and take in a large exhibition in a single visit, our experiences focused them on engaging with fewer exhibits but in different, arguably more focused, ways. The constraint to include only three exhibits in a gift was designed into the *Gift* app from the outset and clearly shaped how participants engaged with the museum, as both givers and receivers engaged closely with just a few selected objects at a time, which led them to make personal interpretations of those objects based on their relationship with their gifting partner.

We saw a strong tendency in the *Gift* app for givers to interpret exhibits through the lens of their relationship to the receiver. Consider for instance the audio message recorded by the teenager Kristin, accompanying a painting in her gift to her mother:

> So, this picture is called Alice in Wonderland, from 1879, and the sofa reminded me a lot of grandma's sofa with the dolls. And the poem says that this is a big sister reading to her little sister, and I think you can imagine me and Leni sitting like this and her reading to me my favourite story.

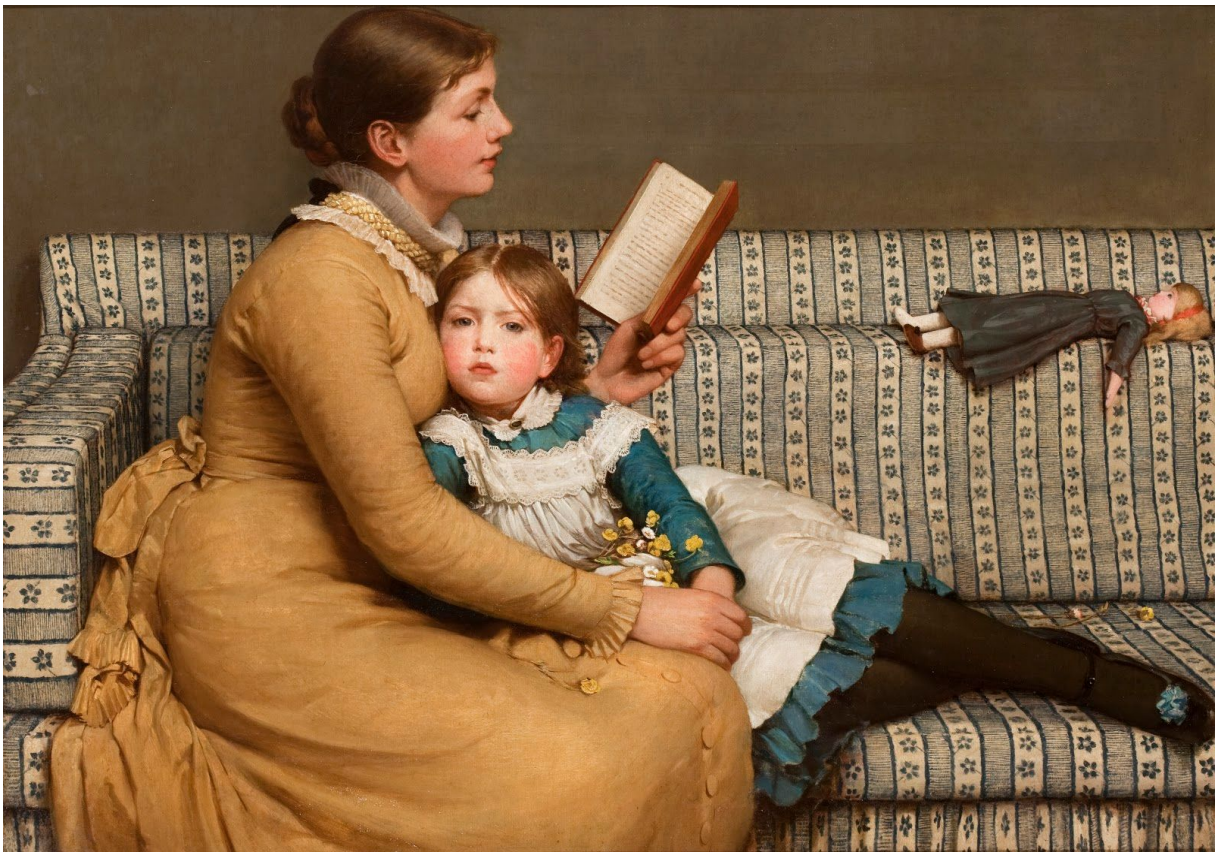

Figure 7: "Alice in Wonderland". Oil painting by George Dunlop Leslie, c1879. Royal Pavilion & Museums, Brighton & Hove (CC-BY-SA).

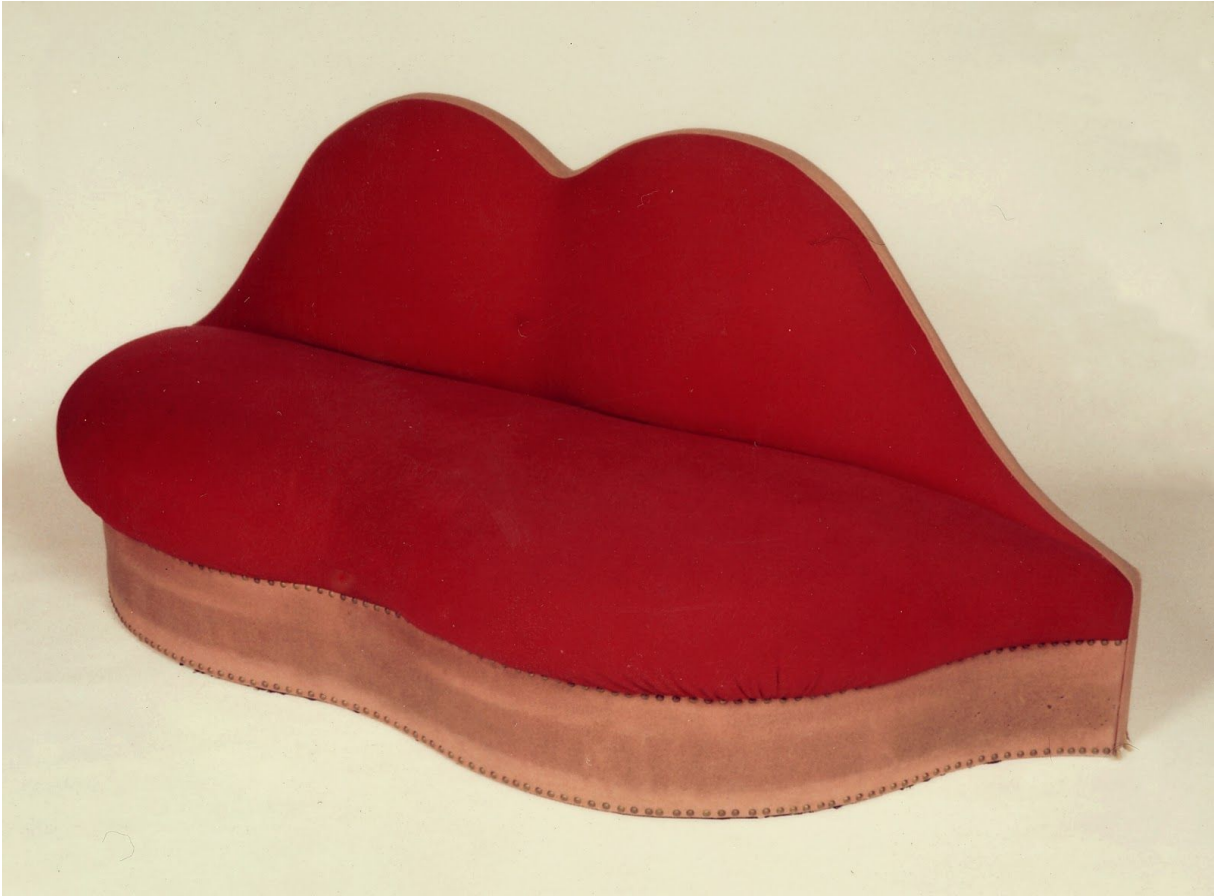

Figure 8: Mae West's Lips sofa, Green & Abbott; Salvador Dali; Edward James, c.1938. Royal Pavilion & Museums, Brighton & Hove (CC-BY-SA).

In the case of Lolly and Jack mentioned earlier, Lolly chooses to give Jack a "lips sofa", because Jack has said that her lips is his favorite thing he likes about her face. These and many other examples show how visitors made personal interpretations that redefined the meanings of exhibits to visitors in terms of their personal relationships and that through this made more intimate connection with the exhibits. From a curator's perspective this might potentially seem crude or deflating; whatever meaning the artwork has according to the artist or museum curators is pushed into the background, and instead it becomes a "prop" in a daughter's reminiscing with her mother or in two teenagers' flirting. Seen from a different perspective, however, these visitors are engaging in personal experiences in which the exhibits play an important part, thus arguably making them come to life in their particular context. This may also involve reappraising the wider historical meaning of the exhibit. Returning to the example of Lolly, she reflected in her interview that:

> My boyfriend is into product design and...furniture and architecture stuff. And that one really, it was bright red, it was the shape of lips, and it really got my attention. It would usually get my attention, but I wouldn't usually read into it. But this is the first time I read into it. So I thought that was interesting, the kind of long drape. But with my boyfriend's lens on, I guess I looked at kind of the artistic side; where they got the inspiration from, the colour, or what material they used to make this and all this other stuff...

Such personal interpretations would seem particularly appropriate in the case of art museums whose exhibits by their very nature would seem open to personal interpretation. However one can argue that they are also valid within other kinds of museum as they introduce new perspectives and voices alongside existing ones, and in a sense acknowledge that many exhibits would have had personal meanings for their original owners that are often lost in the process of a wider historical interpretation.

Although some visitors felt that *Gift* was distracting on some level, and others engaged in offhanded ways that are not likely to be enduringly memorable for them or their receivers, a substantial proportion felt that their increased feelings of closeness to their gifting partner in some way encompassed the gifted object, as well, as discussed above. Many also articulated specific ways in which their engagement with and learning about museum objects increased through their new personal attachments and unusual ways of seeing the gifted objects. The reasons that stood out in our analysis can be grouped into categories of artistic impression, expanded horizons, new motivations, and app mechanics. The first category, artistic impression, cover experiences such as that described by the test participant Neil:

"Some paintings touched me more deeply and on a more personal level having approached them with that person and my connection to them already in mind." Many participants tended to conflate their impressions of the gift with the other person involved (giver or receiver), such as expressed by Sian: "Yes, the gift made me filter my impression thinking of the person the gift was made for. Brought back great memories and important things in our relationship." In terms of expanded horizons, we point to the giver's choices leading receivers to dwell on objects that they would normally ignore, such as Louise's comment that the app "definitely led to focus on a couple of objects more than I would've" and Dan's assertion that they had learned through using *Gift*: "Yes I did, a new way to look at some pieces otherwise I would not have thought about." New motivations also relates to the simple fact that the app directs visitors to objects they might not otherwise have seen, but they discover in the process a new motivation *within themselves* along the way. This can be especially important for givers, who might struggle to change their mental orientation from visiting to gifting. James stated that "I'm not one that can stay in a museum long and will lose interest, but while using the app I stayed pretty interested throughout the museum since I had a 'mission' in a way." Meanwhile, Diana not only maintained but increased her interest:

> Because it makes you engaged. Normally I don't really read those, I just walk and see, Oh, that can be played or whatever. But choosing the gift made me read the caption and like get all the other context in as well, and I was like oh that's interesting. And it like helps make it even deeper. I normally don't read the captions so that was interesting.

Finally, the app's mechanics of requiring voice recordings were cited as reasons for making personal connections to the objects: "Yes, because I had to verbalize why I liked them" (Matthew), and "it's a good way to communicate thoughts that otherwise would go unspoken" (Dan). These interpretations of "intimacy" and "connection" reflect the wide variation in ways that people of all ages and many backgrounds could make sense of the app and sometimes use it to discover unsuspected ways of connecting with and learning about parts of the collection.

Our study of *Never let me go* also showed that participants experienced the exhibited artworks as well as the museum architecture in new and interesting ways. Playing in this way led to fewer but more intimate encounters with the artwork. As Jenny explains, "Maybe we saw less, but some of the things I saw I remember better. Like the shapes I had to enact. Some of these will stay with me much longer because of this experience." And as Nina says, "I definitely think I was looking more into detail than usual. For example, during the explore phase, I was trying to look at things a bit more closely." Somehow the receptive quality of the Avatar role would lead to a different awareness or a specific mindset that allowed for these encounters to become more personal, attentive and intense than usual. As Michael explains, "It felt stimulating. A way of asking new questions. It helps you to use the beginner's mind. To look with fresh eyes on things and step out from your regular thought-inertia". A shift seems to have happened where the role of the observer turned into something else, something more open. Laura describes it in this way:

> I think it was a chance to connect with the art and not just be an observer, but to be part of the paintings but also the whole room. It helped me enjoy it and understand it more. And think about it more. It wasn't just my eyes watching. It was my whole mind observing.

Much focus was also put on emotions, which had a clear effect on the participants' experiences. As Peter explains, "I was more aware of emotions, because I was prompted to be thinking about things I normally don't think about. So yeah, this museum visit was more emotional than my usual museum visits." Thus as with the *Gift* app, the intimacy of the situation, reinforced by the design, would lead players to explore and reflect upon their existing relationship, and it was through this process that an active reinterpretation of the museum context took place. On the negative side, this relational focus would overshadow the museum experience and distract the participants from the curated material. As Lisa says, "I think it became a lot more a tool for the relationship between us rather than the museum itself or the exhibition." On the other hand, it proved to be a powerful tool to engage visitors on a more personal level. As Rebecka explains, "If someone knows how to push your buttons, then the whole thing might even feel like it was curated for you. If it's done in the right sequence, in the right order".

We now drill further into three key design strategies that appear to have been important in mediating this more intimate relationship with the museum: heads-up experience, tone of voice, and vicarious experience.

**Heads-up experience.** Both the *Gift* app and *Never let me go* focus on delivering a heads-up experience in which visitors' attention is directed towards exhibits rather than to the mobile screen. This was done to address previous concerns among museum professionals about screen-based experiences diverting visitors' attention away from the physical exhibits, a concern frequently referred to as "the heads-down phenomenon" (Hsi, 2003; Lyons, 2009; Walter, 1996; Wessel & Mayr, 2007). Delivering a heads-up experience involves extensive use of audio supported by limited amounts of text and interaction on the mobile screen rather than for example video which requires sustained attention to the screen. However, it is also about the "content" of the mobile experience. In our two cases the mobile screen is employed to deliver instructions on how to engage with exhibits in new ways rather than primary interpretation created by the museums' curators. The *Gift* app's developers offloaded all interactions that they reasonably could to audio, with the screen serving more as a support and reassurance than a focus of attention except, of course, when photographing gifted objects and starting audio recordings. Mobile content repeatedly encourages visitors to explore the museum, attend to exhibits and see them in new ways. Any interpretation offered by the museum remains on the walls of the museums, in labels or other audio-visual exhibits, or in guide books, rather than being brought into the mobile. Thus, the mobile experiences are not about alternative ways of providing interpretation, but rather seek to shape how the visitor engages with and thinks about existing resources.

**Tone of voice.** A second important aspect of creating a more intimate relationship with the museum were the distinctive "voices" adopted by the apps when talking to visitors. *Never let me go* created a soundscape with a vocalization of the prompts in a calm and soothing way. Interestingly, it never seemed awkward for the participants to use another person's voice to communicate. As Daria explains, "A voice in a headset is quite intimate for me. And it wasn't her voice, but it was like something that she was saying". The constraints in the communication, in combination with the tone of voice, helped to bring in new perspectives during the museum visit. As Nils points out: "I wouldn't normally ask things like: 'What does it remind you of?'. But here you kind of realize that that's true. You can have different angles where you can come from."

For the *Gift* app, the intimate voice of the narrator described above was critical to the app. It lets the narrator "be really familiar with you because this idea of giving and receiving gifts, and the way she talks about it, is very human. We've all got an experience of it..." (Hunter, 2018, author interview). Hunter's intention was that visitors would perceive the voice as "familiar, relaxed" and "provocative" in the sense of provoking action. He also recognized that this unusual tone was not something that could be achieved in all instances for all people. "Some people really crave that intimacy. Some people [feel] distanced by it because it's just not what they [are] expecting. You have to go one way, and then let people respond" (Hunter, 2018, author interview). Indeed, visitors reported mixed responses to it, with some really tuning into the emotional tone of the experience but others finding it inappropriate or even unnerving. Blast Theory's own reflections on the tone of voice were that:

> There's a tone that's set, and there's the level of familiarity and the language that implies they know you and have a kind of relationship with you even though it's non-personal, and it's suggestive and it leads you through a process of thinking which is intended to be guiding you into a much more reflective space. (Tandavanitj, 2019, author interview)

The participant responses we gained in 2019 seem to indicate that the artists' heightened emphasis on tone of voice had succeeded in establishing the vision described in the quote above. For example, when asked whether they would share the gift they had received with others, Cathy's reply was: "by the way and intensity the woman spoke, I'd think to give the gift only to someone really close to me."

**Vicarious experience.** Both experiences create an intimate experience with the museums and its exhibits by encouraging visitors to see them vicariously through the eyes of another. In *Never let me go* the Controller vicariously experiences the museum through the Avatar. In the *Gift* app the giver does so through the receiver, partly in the form of an imagined experience (i.e, what they will do and how they will feel when they experience the gift), and partly through messages recorded by receivers after receiving their gifts. In this sense, they mirror earlier HCI research on the vicarious experiences of people watching their friends' facial expressions, heart rates, and the like as they rode an amusement ride which led to both parties feeling closer together (Schnädelbach et al., 2008).

However, the question then arises as to how this impacts others who are present in the museum, including "unwitting bystanders" (Benford et al., 2006; Sheridan et al., 2007) who may not be aware of what is taking place. Will they notice and be perturbed by unusual behaviours? In *Never let me go* some participants incorporated other museum visitors into the play to a certain extent. For example, one Avatar started to follow another person instead of the Controller when the prompt was "Follow," and sometimes Controllers would try to make their Avatar do things in front of guards or other visitors to make it more embarrassing and/or fun.

A final point on vicarious experience concerns the extent to which participants might share such intimate experiences. Might gifts be published and shared on social media? At present, this is a matter of choice for the participants. In this context, it is worth noting a subtle but important distinction by "gifting" and "sharing" raised by Spence (2019) who draws on Weiner's concept of inalienability from the gifting literature (Weiner, 1992) to make a separation between the two. Gifts have the ability of a "personal" (or intimate) possession to invoke and symbolize personal memories and knowledge of the giver that cannot otherwise be seen by others and this "inalienable" property makes them distinct from things that are shared more widely without such personal connections. So far, no users of *Never let me go* or the *Gift* app have chosen to share their intimate experiences widely on social media. Museum experience designers should therefore carefully consider this boundary, as well as the broader ethical implications for privacy, before encouraging the sharing of intimate museum experiences. When does intimacy stop and vicarious experience become voyeurism?

**Interpretation**

For our final theme we consider the museum's role in supporting interpersonalized and intimate visitor interpretations and reflect on how this responds to wider changes in the nature and approaches of museums in general. A notable feature of both our experiences is that they encourage visitors to make their own interpretations of exhibits, both for and through others, rather than directly conveying the museum's own interpretation. This requires the museum to step back and make space for interpretation by visitors. This is not only about saying less, but interestingly, also constraining the possibilities to interact with the museum, for example limiting the numbers of exhibits engaged with as discussed above. Making space in this way is a "less is more" strategy; saying less about fewer things makes space for visitors to say more for themselves. Creative practitioners often employ limitations as a tool for scaffolding creativity (Elster, 2000; Mathews, 1997; Rettberg, 2005), and the limitations in the two designs presented here may have a similar effect in reducing the "fear of the blank page." Furthermore, the fact that the designs do not offer any interpretational or educational content, may offer the participants some license to be personal and playful without fear of appearing shallow or uneducated. Limiting the visibility of their interpretations to an interpersonal exchange as discussed earlier further removes the risk of being judged by others than the receiver.

  Both our experiences are therefore "low bandwidth", by which we mean they rely on relatively thin communication channels, at least when compared to media-rich digital tours and immersive experiences. Both experiences reflect the previously mentioned research by

Kaye et al. (2005) where the constrained nature of communication provided space for rich (re)interpretations based on the partners' shared understandings of each other. The tight constraints of *Gift* can also be seen as a rich opportunity to share a contemplative moment that might have been "drowned out" by a more immersive or media-rich experience: "I really do love making a personal connection between my visit and someone I feel will appreciate the gift. Much more intimate than texting a picture from your visit" (Emma from the Gift study). A further benefit of this "thin channel" approach is that both experiences are technically lightweight, requiging little investment in infrastructure or additional content by the museum.

It is also noticeable that both experiences are open and somewhat ambiguous in nature, with visitors being able to interpret what to do in various ways, reflecting the idea that introducing ambiguity into an experience design can be an important strategy for making space for interpretation as discussed by Sengers and Gaver (2006) building on Gaver et al. (2003). In *Gift*, the app's central proposition to users is ambiguous: Inviting them to create gifts out of objects that they cannot buy or own, but just take photos of. It also invites ambiguity/play with what "counts" as an object - e.g. users have included photos of fire extinguishers, selfies etc. Furthermore, the genre of communication is also ambiguous: Are givers creating a personalized guide for the receivers, trying to teach them something, or writing a "postcard", or a personal story, or a joke? In *Never let me go* the instructions are imbued with ambiguity and it is left up to the players to decide how to interpret and act on them. This combination of constraint and ambiguity makes the approaches described here quite different from the dominant tendency of narrative-driven approaches to designing tour guides that emphasize storytelling and rich media or immersive content (cf. Bedford, 2001; Johnson, 2006; Nielsen, 2017; Wong, 2015).

However, the museum is not entirely withdrawn from the experience, but rather provides a scaffold of resources that support visitors in making their own interpretations. These include: the museum environment, exhibits, existing interpretation on the walls, in guidebooks and so forth, and also instructions via digital channels such as *Gift* and *Never let me go*. Instructions are especially important and need to be carefully designed to achieve several goals. As artists with a background in performance and a long history of making interactive digital experiences, Blast Theory brought great expertise in the design of instructions. A previous study of their work *Ulrike and Eamon Compliant* showed how voice and text messages could be skillfully crafted to tell participants where to go, what to do, but also how to behave in public, while also setting an appropriate emotional tone (Tolmie et al., 2012). These same goals are evident in the design of the voice instructions in *Gift* and *Never let me go*, as discussed above. In interactive experiences of these sorts, instructions are the main content, with the skill of the narrator being to guide visitors to tell their own stories. In play, as well as other improvisational practices, a clear framing which helps participants to grasp what is expected of them is also key. In *Never let me go*, the introduction received by the Avatars said:

> Welcome to this Avatar experience. You will soon hear instructions chosen by your partner. Follow these instructions to your own ability and desire. Make it as dramatic or as subtle as you wish. Remember to stay safe and stop whenever you want. When in doubt of what to do, relax and enjoy the art. Now start by doing just that. Enjoy!

This set an overall tone to the experience and helped Avatars to relax by making it clear that it was up to them to interpret the prompts that were sent by the Controller as well as giving them a way out if they needed it.

**Design strategies and challenges**

We have presented two unusual examples of how to deliver personalized experiences to museum visitors, one in which visitors make personal tours as gifts for others, and a second in which one visitor remotely controls the in-the-moment experience of another as they follow them around the museum. Our findings from deploying and studying these in museums reveal that they were generally well received by visitors and that they created opportunities for engaging them in making a particular kind of interpretation in which they view the museums' exhibits through the lens of another person. Underlying these two experiences are two important design strategies, *interpersonalization* and *intimacy*.

The strategy of designing for interpersonalization differs from previous approaches to personalization and customization in two important ways. First, conventional approaches focus on bi-partite interactions between the "business" (in our case the museum) and the individual "consumer" (the visitor). Interpersonalization, on the other hand, involves a tripartite relationship among two "consumers" – the giver/controller and receiver/avatar – with the "business" or museum supporting and scaffolding the relationship. Second, personalization has largely been seen as an algorithmic process and customization a more human-driven one, whereas interpersonalization sits between the two, being primarily driven by humans who do the heavy lifting of tailoring experiences, but scaffolded by the system that provides the instructions and resources to support them. Thus, another way of phrasing our discussion about interpersonalization in the context of the museum is that there are now two kinds of visitor in the picture; one who receives the personalized experiences, with some context provided by the museum, but also a second who co-creates the experience with the museum. The emergence of the co-creator is especially interesting as our examples suggest that people can enjoy and benefit from interpersonalizing experiences for others. In other words, rather than being a chore or hard work, there may be opportunities to engage people who wish to undertake the work of interpersonalizing for others, perhaps because it demonstrates their positive feelings towards the receiver (as discussed in Spence et al., 2019) and/or because it is entertaining or informative in its own right. We also note the possibility to combine our approach to interpersonalization with more algorithmic approaches in future work, for example using algorithms to recommend potential exhibits of interest or even learning from how humans interpersonalize experiences to develop more subtle algorithms.

Our second strategy of designing for intimacy similarly involves adopting a tri-partite rather than bi-partite perspective. The intimacy here is not only between pairs of visitors, but also with the museum and its exhibits. By fostering intimacy between visitors, the museum may then open up an opportunity to create more intimate relations with its own exhibits. This intimacy arises from deep personal knowledge of the other person as required to choose the right gift or instruction for them, but also to a degree on a vicarious experience, being able to see (or at least imagine) their experience. It is also scaffolded through the careful design of instructions including tone of voice, which may need to differ from that normally adopted by the business (e.g., a voice that is unlike a conventional tour guide or curator). Our findings also speak to the design of vicarious experiences as previously considered in HCI. *Never let me go* delivers an overtly vicarious experience, while *Gift* perhaps relies more on the giver's imagination of how any object might be selected, described, and received, and the receiver's imagination of the gift-creation process. Like the vicarious experiences on roller coasters discussed above (Schnädelbach et al., 2008), there is also an element of carefully managing discomfort both within the design and as enacted by all participants, not only in terms of the potential embarrassment of a poorly chosen experience, but also considering the presence and impact on other bystanders in the museum. However, the risk of "making a fool of oneself" as well as having a hidden purpose that excludes other visitors are both components that strengthen the intimacy of the experience to begin with.

Having set out these two overarching strategies, we reflect on opportunities for realizing them in practice:

- Personalization can be about getting visitors to personalize for each other, rather than the museum doing the personalisation – manually or algorithmically. This can be a low-cost strategy, requiring little technical infrastructure beyond visitors' own devices and little new digital content, as visitors make this for each other.
- Interpersonalization may foster a new kind of interpretation, one in which someone interprets an exhibit for another, rather than the museum interpreting it for them, or them interpreting it for themselves. This can be seen as opening up a space of "second-person" interpretations to complement the long-dominant third-person perspective of the museum's canonical interpretations and the more recent emergence of the first-person perspective of the individual visitors' interpretations.
- Employing intimacy by getting visitors to bring their personal relationships into the museum or to have a vicarious experience can be a powerful approach, but needs to be treated with caution lest it backfire, leading to overly uncomfortable or intrusive experiences. It may require the museum to adopt a different voice than usual, but perhaps then there are also risks about authenticity and appropriateness. These are important questions that require further research.
- Interpersonalization and intimacy are subtle strategies that require the museum to be prepared to stand back and hand over control to visitors. Consequently, it may lose

- control of their interpretations which, for example, might become more about visitors' own stories. It may also lose control of the technologies as both of our experiences could potentially be deployed in a museum without its direct involvement as we discuss further below.
- We note that there may be interesting possibilities to extend these approaches to remote visiting situations, something that may come to the fore given the consequences of the COVID-19 global pandemic in the short term (mid-2020 at the time of writing) and the pressures of climate change in the long term, though this is not a topic we have directly addressed in this paper.

However, we also call attention to potential challenges arising from our two strategies. Most notable among these is how to reintroduce the museum's perspective back into the experience. How do these strategies stack up against the wider educational agenda of many museums, either formally through school outreach programmes, or informally through championing participatory perspectives and ideologies, for example around diversity and inclusion? How are these to be brought into the picture? How do museums still interject their values and knowledge into the dialogue? We suggest some ways in which museums might respond to these challenges:.

- Museums are still responsible for choosing which exhibits are available to visitors, where, and when. They can still provide conventional interpretations outside of the immediate digital experience, through labels, for example. They can also use any other digital or analogue means of exposing visitors to "the facts" as the curators understand them, and to encourage as much or as little structured dialogue with – or challenge to – the curatorial stance as they wish.

- Museums can offer links to official interpretations during a digital experience such as the *Gift* app or *Never let me go* that invites personal interpretations. We suggest that one possibility is to allow visitors to access these officially curated interpretations afterwards, building on the approach of Fosh et al. (2013) in which visitors to a sculpture garden first engaged with each sculpture in an experiential and open way before being offered "official" information as they walked away.
- There are opportunities to customize the experience, including branding, initial message and instructions. Interestingly, both of our experiences are sufficiently generic that they might in principle be rolled out in nearly any museum. However, as Blast Theory learned in adapting *Gift* for the Munch Museum, a blanket approach may be easier said than done. There may easily be issues of language, tone, policy, and infrastructure that will shape the uptake and content of the app, which may in turn shape common visitor experiences. For example, human resource constraints and museum priorities precluded Munch Museum staff from engaging with individual visitors personally about the app, which instead appeared as a free offering on its official price list. Both museums advertised the app using beautifully designed postcards placed inside the museum, yet these led directly to relatively few new users regardless of context – in cluttered competition with many other such cards in Brighton, or prominently and exclusively displayed on the gallery walls in Oslo. *Gift* referred to "objects" for the eclectic, Victorian-era collections of the Brighton Museum, but this term made little sense in the Munch museum which is devoted entirely to visual arts. Similar re-evaluations and adjustments have been made in subsequent deployments.

- Museums can also provide example experiences of unusual apps such as the ones discussed here to illustrate the process, set expectations, and engage visitors: for example, the *Gift* app as deployed at Brighton Museum in 2019 began with the chance to receive a gift made by the museum's curators. This approach might extend to drawing on the voices of "friends" of the museum and other influencers, thereby deepening the sense of investment in "their" museum.
- Finally, we stress that the kinds of approaches that we propose here are not intended to be exclusive or even to replace other ways of engaging with the museum, but rather should be seen as complementary, engaging visitors in new ways that might lead to or follow on from other forms of experience. A challenge for future work is to better understand how to connect them to these other existing kinds of experience.

Although there are many points of the visitor journey where museums can interject their voices, the kinds of experience we have presented here undoubtedly do involve a shift in the balance of control between museums and visitors, reflecting the longstanding trend in the new museology literature. This raises further challenges associated with giving up control to visitors. What happens if they make uninformed interpretations or say terrible things to each other, for example involving hate speech or bullying? This is another argument for restricting the visibility of visitors' interpretations to themselves rather than placing them on social media sharing platforms. On the other hand there are opportunities to learn from visitors too. What exhibits do they choose and why and how might this inform future curatorial choices?

Such questions hinge on the question of ownership. First, who owns the interpretations that visitors generate, especially given that they may be highly personal and sensitive? What level of analytics or moderation should museums undertake? Second, related to this, who owns the museum and its exhibits? It is perhaps a sobering thought that the two experiences we have introduced are potentially relocatable to many museums without them needing to be involved at all. They require no heavy content development or infrastructure that might not already be publicly available. Indeed, these designs invite us to see the screens of visitors' own devices as their own personal territory, under their control, just as the walls of the museum are likely to remain under its control.

Related to ownership is the question of appropriation. HCI has tended to view appropriation as a positive aspect of users' engagement with digital technologies in which they adapt interfaces to their own (sometimes unexpected) purposes. However, the term has quite a different connotation in museums, where it typically refers to borrowing or even stealing another culture's artefacts and histories without permission, a deeply problematic challenge in the era of post-colonialism (Ziff & Rao, 1997). Our approach raises the question of whether enabling visitors to directly appropriate museum exhibits for their own personal purposes, such as making gifts for others, ppens up the museum to new and different voices and might even allow people to reclaim their historical artefacts, or alternatively whether it runs the risk of extending the mis-appropriation of others' artefacts from an institutional to a personal level.

Answering such challenging questions falls outside the scope of the data that we gathered and hence of this current paper. However we note that inviting visitors to make personal interpretations on their own devices will inevitably lead to these kinds of tensions and questions and that exploring them in practice is a key goal for future research into interpersonalized and intimate experiences.

**Conclusions**

We have presented the design and deployment of two museum visiting apps that involve visitors creating experiences for one another, one by transforming exhibits into personal gifts, and the other by having one visitor direct a partner's actions in real-time as they explore the museum. While unusual by conventional standards, especially when compared to established virtual tour guides, we argue that both were successful at creating engaging and thought-provoking experiences that led visitors to see the museum and its exhibits - and perhaps each other too - in new ways.

Our reflections on these experiences informed two overarching design strategies for designing museum experiences: interpersonalization in which visitors personalize experiences for each other; and intimacy in which such experiences draw on and reinforce more "close up and personal" associations, both among visitors and between visitors and exhibits. We further reflected on how these two strategies raise new opportunities and challenges for museum designers, especially how they invite museums to take the brave step of standing back to make space for visitors to generate their own interpretations, while still providing the resources to underpin these and trying to shape them more generally.

As a final comment, while our focus has been museums, we note that the approach of interpersonalizing intimate experiences might extend to other domains. Other cultural experiences such as games and entertainment are examples with clear parallels to interactive experiences in cultural heritage institutions. More commercial kinds of gifting provides another obvious example. Social media and personal communications may also benefit from this more nuanced approach to the sharing of experiences in public places. We also note the potential to impact on the design of learning experiences, especially within the museum context but also in almost any other context that can leverage the ability to create interpersonalized, intimate experiences from unusual external stimuli.

**Acknowledgments**

The research reported in this article formed part of the GIFT project ([gifting.digital](gifting.digital)), which has received funding from the European Union's Horizon 2020 research and innovation programme under grant agreement No 727040.

**Author bios**

**Karin Ryding:** PhD Fellow, Digital Design Department, IT University of Copenhagen. Karin's research interest includes critical play, game design, and cultural heritage. She holds an MA in Visual Culture Studies and a BA in Game Design. Outside of her research career, Karin has worked as a professional game designer.

**Jocelyn Spence:** Research Fellow, Mixed Reality Lab, University of Nottingham. Jocelyn is an HCI and design researcher working at the intersection of performance and human-technological interaction, particularly the intensification of conversational storytelling. Her work has come to include focuses on gifting and museum-based interactions aiming for meaningful engagement. She also works in design fiction, ethics, and theory.

**Anders Sundnes Løvlie:** Associate Professor, Digital Design Department, IT University of Copenhagen. Anders does research on the intersection of design research and media studies, focusing in particular on experience design, locative media and play. He was the coordinator for [the GIFT project](#) and leads the research group on [Media, Art and Design](#) (MAD).

**Steve Benford:** Dunford Professor of Computer Science, Mixed Reality Lab, University of Nottingham. Steve's research explores how digital technologies, and foundational concepts and methods to underpin these, can support cultural and creative experiences including new forms of museum performance. He is currently directing the Horizon Centre for Doctoral Training.